\documentclass[psfig,usenatbib,useAMS,usegraphicx,subfig]{mn2e}
\usepackage{amssymb}
\usepackage{times}
\begin{document}

\title[Debris disc stirring]
  {Debris disc stirring by secular perturbations from giant planets}

\author[A. J. Mustill \& M. C. Wyatt]
  {Alexander J. Mustill$^1$\thanks{Email: ajm233@ast.cam.ac.uk},
   Mark C. Wyatt$^1$ \\
  $^1$ Institute of Astronomy, University of Cambridge, Madingley Road,
  Cambridge CB3 0HA, UK}

\maketitle

\begin{abstract}
Detectable debris discs are thought to require dynamical excitation (`stirring'), so that planetesimal collisions release large quantities of dust. We investigate the effects of the secular perturbations of a planet, which may lie at a significant distance from the planetesimal disc, to see if these perturbations can stir the disc, and if so over what time-scale. The secular perturbations cause orbits at different semi-major axes to precess at different rates, and after some time $t_{\mathrm{cross}}$ initially non-intersecting orbits begin to cross. We show that $t_{\mathrm{cross}}\propto a_{\mathrm{disc}}^{9/2}/(m_{\mathrm{pl}}e_{\mathrm{pl}}a_{\mathrm{pl}}^3)$, where $m_{\mathrm{pl}}$, $e_{\mathrm{pl}}$ and $a_{\mathrm{pl}}$ are the mass, eccentricity, and semi-major axis of the planet, and $a_{\mathrm{disc}}$ is the semi-major axis of the disc. This time-scale can be faster than that for the growth of planetesimals to Pluto's size within the outer disc. We also calculate the magnitude of the relative velocities induced amongst planetesimals and infer that a planet's perturbations can typically cause destructive collisions out to 100's of AU. Recently formed planets can thus have a significant impact on planet formation in the outer disc which may be curtailed by the formation of giant planets much closer to the star. The presence of an observed debris disc does not require the presence of Pluto-sized objects within it, since it can also have been stirred by a planet not in the disc. For the star $\epsilon$~Eridani, we find that the known radial velocity planet can excite the planetesimal belt at 60\,AU sufficiently to cause destructive collisions of bodies up to 100\,km in size, on a time-scale of 40\,Myr.
\end{abstract}

\begin{keywords}
  circumstellar matter --
  planetary systems: formation --
  planetary systems: protoplanetary discs --
  stars: individual: $\epsilon$~Eridani --
  stars: individual: Fomalhaut.
\end{keywords}

\section{Introduction}

Since the first detections in thermal infrared \citep{1984ApJ...278L..23A} and scattered light \citep{1984Sci...226.1421S}, it has become clear that many main-sequence stars are surrounded by discs of dust grains. The dust grains have only a short lifetime compared to the age of the star, being ground down in collisional processes until they are small enough to be ejected from the system by radiation pressure \citep{2003ApJ...598..626D,2005AandA...433.1007W}. The existence of a dusty debris disc therefore implies a large reservior of planetesimals, parent bodies with longer collisional lifetimes whose collisions are producing the observed dust. Such planetesimals must be colliding with sufficient relative velocity to produce enough dust to be observable, and so the disc must have some degree of dynamical excitation, with non-zero eccentricities and inclinations causing high relative velocities. Although the formation of planetesimals is still not understood, it is usually assumed that they form on coplanar, near-circular orbits in the protoplanetary disc \citep[e.g.,][]{2008ApJS..179..451K}. Therefore, there must be some means of exciting the planetesimals' eccentricities for the disc to become visible. This is referred to as \emph{stirring.}

The origin of debris disc stirring is not yet known and has yet to be thoroughly investigated. The most comprehensive model of this is $self-stirring$ \citep[and references therein]{2008ApJS..179..451K} where a planetesimal disc evolves due to mutual low-velocity collisions resulting in the growth of the planetesimals. In this model the disc is stirred, and collisions become destructive, when the largest planetesimals reach Pluto's size, and their gravitational perturbations excite the remaining smaller bodies.

Such planet formation models, however, ignore the effect of the formation of massive planets on the evolution of material. Yet we know that large numbers of massive planets exist, including within debris disc systems \citep[e.g.,][]{2006AJ....132.2206B,2008Sci...322.1345K}. The existence of a massive planet in a disc system can have several consequences, including migration of the planet \citep{1984Icar...58..109F} and scattering of embryos from the inner disc near the planet to the outer disc \citep*{2004ApJ...614..497G}. The formation of a planet may also be able to speed planet formation exterior to the planet when the gas disc is still present \citep{2005ApJ...626.1033T}. In this paper, we consider the effects of a planet's gravitational secular perturbations on the disc, showing that they can stir the disc on timescales of typically several 10s of Myr.

This paper is organised as follows. In \S2 we review the secular theory we are using to model the effect of the planet on the disc. In \S3 we derive an analytical estimate for the time the planet takes to stir the disc. In \S4 we investigate the relative velocity distribution imposed by the planet. In \S5 we discuss the model we use to ascertain the outcomes of collisions. In \S6 we discuss the implications of planet stirring for debris disc evolution. Finally, in \S7 we summarise the paper.

\section{Dynamics} \label{sec:dynamics}

We consider the orbital evolution of planetesimals, treated as massless test particles, under the perturbing influence of $N$ massive planets. We shall consider only secular perturbations. In the following, unsubscripted orbital elements refer to test particles and subscripted to the planet(s).

We use classical Laplace--Lagrange secular theory to model the long-term orbital evolution. This is valid if planetesimals are not near mean motion resonance and eccentricities are low. The planetesimal's complex eccentricity $z=e\exp\left(\mathrm{i}\varpi\right)$ can be decomposed into a forced component $z_{\mathrm{f}}$ and a proper component $z_{\mathrm{p}}$:
\begin{eqnarray}\label{eq:z}
z(t) &=& z_{\mathrm{f}}(t)+z_{\mathrm{p}}(t)\\
     &=& -\sum\limits_{i,j=1}^N\frac{A_je_{ji}}{A-g_i}\exp\left[\mathrm{i}\left(g_it+\beta_i\right)\right] + e_{\mathrm{p}}\exp\left[\mathrm{i}\left(At+\beta\right)\right]\nonumber
\end{eqnarray}
\citep{1999ApJ...527..918W}. Here, $g_i$ and $e_{ji}$ are the eigenvalues and eigenvector components of the solution for the planets, independent of the planetesimal's location, and $\beta_i$, $\beta$, and $e_{\mathrm{p}}$ are constants of integration; $e_{\mathrm{p}}$ is the proper eccentricity. The constants $A_j$ are given by
\begin{equation}\label{eq:aj}
A_j=-\sqrt{\mathcal{G}m_\star /a^3}\frac{m_j}{m_\star} \alpha_j \bar{\alpha}_j b^{(2)}_{3/2}(\alpha_j)/4,
\end{equation}
with $m_j$ the mass of the $j$-th planet, and $A$ is given by $A=\sum_{j=1}^N B_j$, with
\begin{equation}\label{eq:bj}
B_j=+\sqrt{\mathcal{G}m_\star /a^3}\frac{m_j}{m_\star} \alpha_j \bar{\alpha}_j b^{(1)}_{3/2}(\alpha_j)/4,
\end{equation}
$\alpha_j=a_j/a$ for an interior planet and $\alpha_j=a/a_j$ for an exterior planet, $b^{(s)}_j(\alpha)$ are the Laplace coefficients (see \citealt{1999ssd..book.....M}), and $\bar{\alpha}_j=1$ for an interior planet and $\bar{\alpha}_j=\alpha_j$ for an exterior planet. 

The behaviour of the complex eccentricity is to precess at a rate $A$ on a circle of radius $e_{\mathrm{p}}$ about the forced complex eccentricity $z_{\mathrm{f}}$. For more than one planet, $z_{\mathrm{f}}$ is itself evolving in time. If there is only one planet, its orbit is fixed, and so is $z_{\mathrm{f}}$.

Henceforth we shall concentrate on the case of a single planet. We shall return briefly to the multi-planet case in \S\ref{s:multi}. Denoting the planet's elements now by $_{\mathrm{pl}}$, we have
\begin{equation}\label{eq:sec_soln}
z(t)=\frac{b^{(2)}_{3/2}}{b^{(1)}_{3/2}}e_{\mathrm{pl}}\exp\left(\mathrm{i}\varpi_{\mathrm{pl}}\right)+e_{\mathrm{p}}\exp\left[\mathrm{i}\left(At+\beta\right)\right].
\end{equation}
We shall later concentrate on the case where the planetesimals' orbits are initially circular. In this case, $e_{\mathrm{p}}=e_{\mathrm{f}}$ and $\beta=-\varpi_{\mathrm{pl}}$.

Using the leading-order approximations to the Laplace coefficients \citep{1999ssd..book.....M}:
\begin{equation}\label{eq:Laplace expansion}
b^{(1)}_{3/2}\sim3\alpha, \qquad b^{(2)}_{3/2}\sim\frac{15}{4}\alpha^2 \qquad \mathrm{as\,\,} \alpha \to 0,
\end{equation}
we have
\begin{equation}\label{eq:ef approx}
e_{\mathrm{f}}\sim \frac{5}{4}\alpha e_{\mathrm{pl}},
\end{equation}
and
\begin{equation}\label{eq:A approx}
A\sim \sqrt{\mathcal{G}m_\star/a^3}\frac{3}{4}\mu\alpha^2\bar{\alpha},
\end{equation}
in the limit of small $\alpha$.

\begin{figure}
  \includegraphics[width=.5\textwidth]{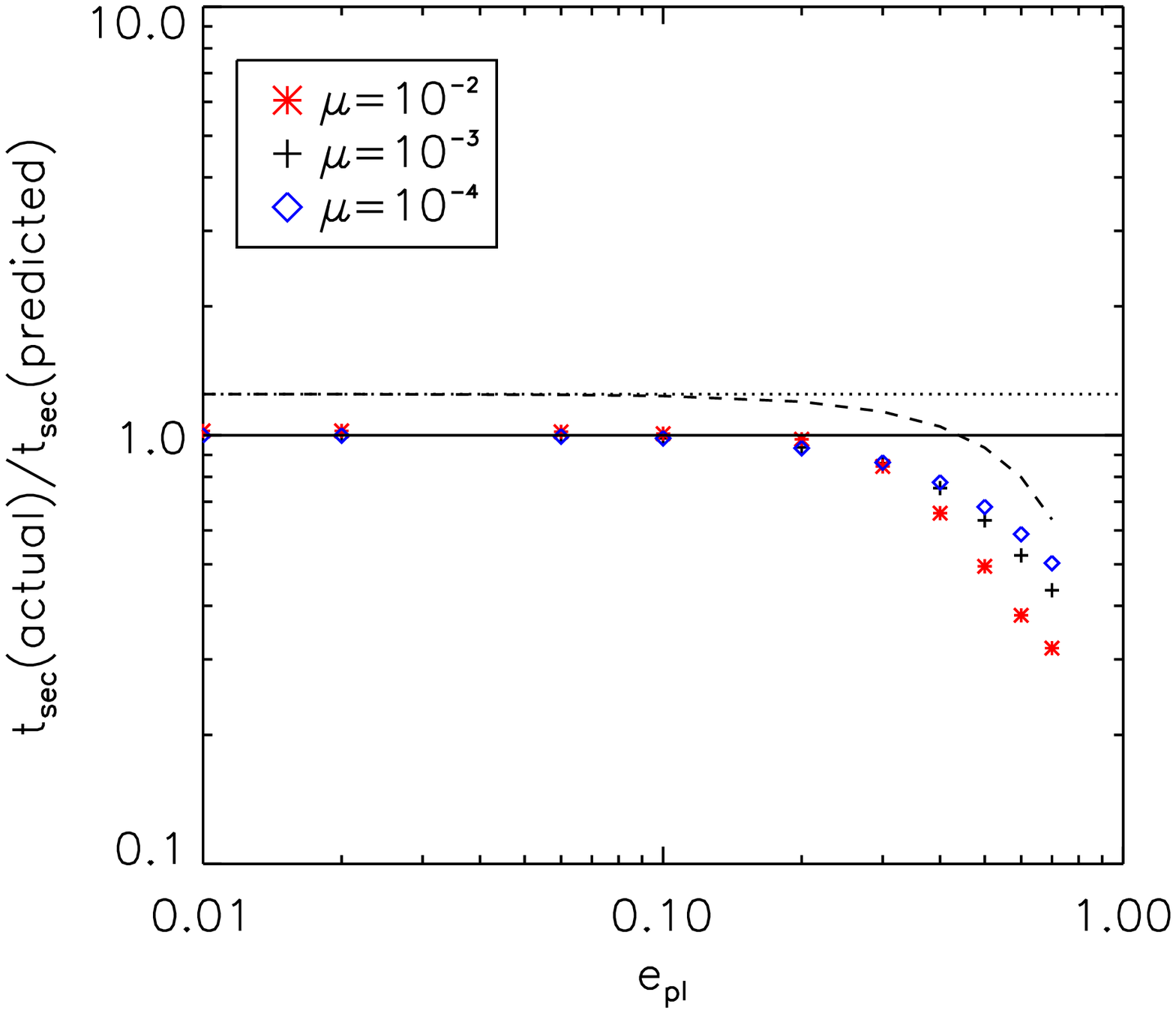}
  \includegraphics[width=.5\textwidth]{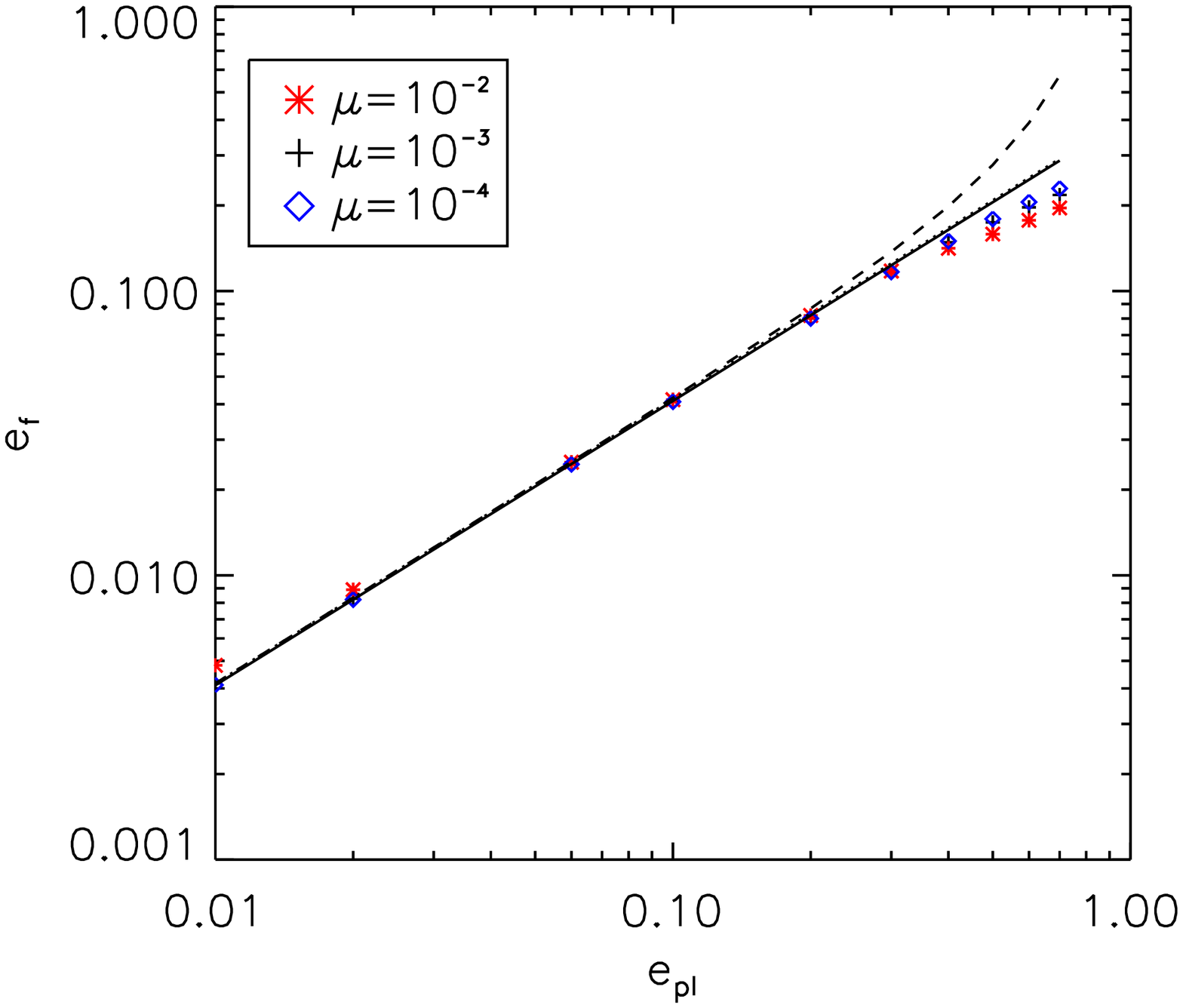}
  \caption{Secular precession time-scale $t_{\mathrm{sec}}=2\pi/A$ (top) and forced eccentricity $e_{\mathrm{f}}$ (bottom) for a planetesimal on an initially circular orbit at $a=15$\,AU, with $a_{\mathrm{pl}}=5$\,AU, $m_\star=1\mathrm{M}_\odot$, and $\mu=m_{\mathrm{pl}}/m_\star$ and $e_{\mathrm{pl}}$ varying as shown. Symbols: N-body simulations using RADAU \citep{1985dcto.proc..185E}. Solid line: analytical prediction from Laplace--Lagrange theory. Dotted line: analytical prediction from Laplace--Lagrange theory with Laplace coefficients approximated to leading order in $\alpha$ (Equations \ref{eq:Laplace expansion}). Dashed line: analytical prediction from Heppenheimer-Kaula theory.}
  \label{fig:tsec,ef}
\end{figure}

In Figure~(\ref{fig:tsec,ef}) we compare $t_{\mathrm{sec}}$ as predicted by Laplace--Lagrange theory to that obtained by numerical integration (using RADAU, \citealt{1985dcto.proc..185E}). Precession timescale $t_{\mathrm{sec}}$ was calculated by fitting a sinusoid to the eccentricity output from the integration\footnote{Orbital elements from the simulations are derived from canonical Jacobi coordinates (motion of planet referred to the star, motion of test particle referred to the star-planet barycentre). This eliminates variations in eccentricity (of order $4\mu/\sqrt\alpha$) on the Keplerian timescale \citep{2003ApJ...592.1201L}, which exceed the forced secular eccentricity at large distances from the star. Use of Jacobi coordinates also ensures no dependence of secular behaviour on initial mean longitude at the same semi-major axis, which is not the case for heliocentric coordinates.}. While Laplace--Lagrange theory is very accurate at small planetary eccentricities, it overestimates the precession time-scale and forced eccentricity at high $e_{\mathrm{pl}}$. For high eccentricity planets, we therefore adopt the theory of \cite{1978A&A....65..421H} based on the disturbing function expansion of \cite{1962AJ.....67..300K}. This is an expansion in $\alpha$ with no restriction on $e_{\mathrm{pl}}$, in contrast to the Laplace--Lagrange expansion which is for small $e_{\mathrm{pl}}$ without restriction on $\alpha$ (provided there is no mean motion resonance). This predicts qualitatively the same behaviour, with forced eccentricity
\begin{equation}\label{eq:ef H}
e_{\mathrm{f}}=\frac{5\alpha e_{\mathrm{pl}}}{4\left(1-e_{\mathrm{pl}}^2\right)},
\end{equation}
and precession rate
\begin{equation}\label{eq:A H}
A=\sqrt{\mathcal{G}m_\star/a^3}\frac{3\mu\alpha^2\bar{\alpha}}{4\left(1-e_{\mathrm{pl}}^2\right)^{3/2}}.
\end{equation}

From Figure~(\ref{fig:tsec,ef}), we see that, while the Heppenheimer-Kaula theory provides a better description of secular behaviour of $t_{\mathrm{sec}}$ than Laplace--Lagrange theory for high $e_{\mathrm{pl}}$, it actually gives a worse fit for $e_{\mathrm{f}}$. We therefore omit the factor $\left(1-e_{\mathrm{pl}}^2\right)^{-1}$ from $e_{\mathrm{f}}$ from now on, as an empirical correction. We also see in Figure~(\ref{fig:tsec,ef}) that the Heppenheimer-Kaula solution is overestimating $t_{\mathrm{sec}}$ for low $e_{\mathrm{pl}}$. This is due to higher order terms in $\alpha$ which the Heppenheimer-Kaula solution does not take into account. For comparison, the Laplace--Lagrange solution is also plotted to leading order in $\alpha$. As $\alpha$ becomes small, the Heppenheimer-Kaula solution performs better at small $e_{\mathrm{pl}}$. Note that at high $e_{\mathrm{pl}}$ there is also a dependence of $t_{\mathrm{sec}}$ and $e_{\mathrm{f}}$ on $\mu$ beyond that predicted by either theory; this is probably due to both theories being based on treating the perturbations only to leading order in $\mu$.

\begin{figure}
  \includegraphics[width=.5\textwidth]{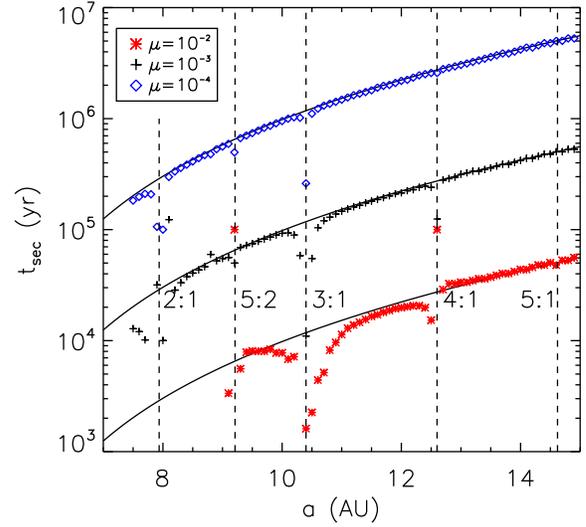}
  \caption{Secular precession time-scale $t_{\mathrm{sec}}$ as a function of planetesimal's semi-major axis, with $a_{\mathrm{pl}}=5$\,AU, $m_\star=1\mathrm{M}_\odot$, $\mu=0.001$, and $e_{\mathrm{pl}}=0.1$. Solid line: L--L prediction. Crosses: simulation results. Note the presence of strong mean motion resonances at the indicated locations. Planetesimals which were ejected within the integration time ($10^5$, $10^6$, and $10^7$ years for $\mu=10^{-2}$, $\mu=10^{-3}$, and $\mu=10^{-4}$ respectively), mostly close to the 2:1 resonance with $\mu=10^{-2}$, are not shown. The analysis in this paper is not valid for planetesimals in resonance.}
  \label{fig:tsec-a}
\end{figure}

Figure~(\ref{fig:tsec-a}) shows the dependence of $t_{\mathrm{sec}}$ on $a$ for three different planet masses. The Laplace--Lagrange theory describes well the behaviour of test particles, except for those near mean motion resonance with the planet. The effect of resonances covers a larger range of semi-major axes as planetary mass is increased. The analysis we present below is based on a purely secular theory and so does not apply for planetesimals near resonance. For $\mu=10^{-3}$ and $a_{\mathrm{pl}}=5$\,AU, this excludes planetesimals within 8.5\,AU and those and those at 10--11\,AU.

In subsequent sections we shall derive expressions for the relative velocities in the disc and the time-scale for disc stirring correct to first order in $\alpha$. Henceforth we use Equation~(\ref{eq:A H}) for the precession rate, and Equation~(\ref{eq:ef approx}) for the forced eccentricity.

The secular precession rate $A$ is a strong function of $a$ (Equation \ref{eq:A H}, Figure \ref{fig:tsec-a}). Furthermore, the forced eccentricity $e_{\mathrm{f}}$ also depends on $a$ (Equation~\ref{eq:ef approx}). The eccentricity evolution and apsidal precession of planetesimals at different semi-major axes is different, and this differential precession raises the possibility that the initially non-intersecting orbits may cross after a certain time $t_{\mathrm{cross}}$. We investigate this possibility in the next section.

\section{Time-scale for orbit crossing} \label{s:timescale}

We are interested in whether a planet's secular perturbations can cause neighbouring planetesimals to collide. Consider two planetesimals at semi-major axes $a_1$ and $a_2$, with $a_1 < a_2$. The orbits are interecting at a time $t$ if $r_1\left(a_1,e_1(t),\varpi_1(t),\theta\right) \ge r_2\left(a_2,e_2(t),\varpi_2(t),\theta\right)$ for some true longitude $\theta$. Writing the equation of an ellipse to first order in eccentricity, this condition becomes
\begin{equation}\label{eq:r1>r2}
a_1\left[1-e_1\cos(\varpi_1-\theta)\right] \ge a_2\left[1-e_2\cos(\varpi_2-\theta)\right]
\end{equation}
Now, we consider closely separated orbits, so that $a_2 = a_1 + \mathrm{d}a$. We then require
\begin{eqnarray}
0 & \le & 1 + e_1\cos\left(\varpi_1-\theta\right) 
+ a_1\cos\left(\varpi_{\mathrm{pl}}-\theta\right)\left(\frac{\upartial e_{\mathrm{f}}}{\upartial a}\right)_t\nonumber\\
& + & a_1\cos\left(\varpi_1-\theta\right)\frac{\mathrm{d}e_{\mathrm{p}}}{\mathrm{d}a}
- a_1e_{\mathrm{p,1}}t\sin\left(\varpi_1-\theta\right)\frac{\mathrm{d}A}{\mathrm{d}a}\nonumber\\
& - & a_1e_{\mathrm{p,1}}\sin\left(\varpi_1-\theta\right)\frac{\mathrm{d}\beta}{\mathrm{d}a}+ \mathcal{O}\left(e_1^2\right).\label{eq:dr/da<0}
\end{eqnarray}
All the terms are $\mathcal{O}(e)$ except for the first term and the term $- a_1e_{\mathrm{p,1}}t\sin\left(\varpi_1-\theta\right)\frac{\mathrm{d}A}{\mathrm{d}a}$ which becomes order unity for sufficiently large time $t$, provided that $\frac{\mathrm{d}A}{\mathrm{d}a} \ne 0$ (this condition is satisfied for single-planet systems, but multi-planet systems can have $\frac{\mathrm{d}A}{\mathrm{d}a} = 0$; see \S\ref{s:multi}). If $\frac{\mathrm{d}A}{\mathrm{d}a} = 0$ then second-order terms in the Taylor series must be considered; the term of order unity will be $\propto e_{\mathrm{p,1}}{d}a\mathrm{\,} t^2\frac{\mathrm{d}^2A}{\mathrm{d}a^2}$.

We now use the leading-order expansions for the Laplace coefficients (Equation \ref{eq:Laplace expansion}), so with small eccentricity $e$ and planet/planetesimal semi-major axis ratio $\alpha$ we have
\begin{equation}\label{eq:dr/da}
0 \ge 1+bAte_{\mathrm{p}}\sin\left(At+\beta-\theta\right)+\mathcal{O}\left(e\right)
\end{equation}
for orbit intersection, where $b=7/2$ for an internal perturber and $b=-3/2$ for an external perturber. The second term is of order unity when $t$ is of order $e_{\mathrm{p}}^{-1}$.

At the crossing time $t=t_{\mathrm{cross}}$,
\begin{equation}\label{eq:dr/da=0}
1+bAt_{\mathrm{cross}}e_{\mathrm{p}}\sin\left(At_{\mathrm{cross}}+\beta-\theta\right)+\mathcal{O}\left(e_{\mathrm{p}}^2\right)=0.
\end{equation}
Now, $|\sin\left(At_{\mathrm{cross}}+\beta-\theta\right)|\le 1$, and it will attain its maximum for some $\theta$, so
\begin{equation}\label{eq:tcross}
t_{\mathrm{cross}}\sim \frac{1}{|b|Ae_{\mathrm{p}}}.
\end{equation}
Note that this has no explicit dependence on $e_{\mathrm{f}}$, but can have a dependence on $e_{\mathrm{f}}$ if $e_{\mathrm{p}}$ depends on $e_{\mathrm{f}}$. In the remainder of this paper, unless otherwise stated, we assume the orbits are initially circular; this means that $e_{\mathrm{f}}=e_{\mathrm{p}}$. Then, for an internal perturber, we have
\begin{eqnarray}\label{eq:tcrossint}
t_{\mathrm{cross}}&\sim& 1.53\times 10^3
\frac{\left(1-e_{\mathrm{pl}}^2\right)^{3/2}}{e_{\mathrm{pl}}}
\left(\frac{a}{10\mathrm{\,AU}}\right)^{9/2}\nonumber
\\ &&
\times
\left(\frac{m_\star}{\mathrm{M}_\odot}\right)^{1/2}
\left(\frac{m_{\mathrm{pl}}}{\mathrm{M}_\odot}\right)^{-1}
\left(\frac{a_{\mathrm{pl}}}{1\mathrm{AU}}\right)^{-3}\mathrm{yr},
\end{eqnarray}
while for an external perturber, we have
\begin{eqnarray}\label{eq:tcrossext}
t_{\mathrm{cross}}&\sim& 1.11\times 10^3
\frac{\left(1-e_{\mathrm{pl}}^2\right)^{3/2}}{e_{\mathrm{pl}}}
\left(\frac{a_{\mathrm{pl}}}{10\mathrm{AU}}\right)^4 \nonumber
\\ && 
\times
\left(\frac{m_\star}{\mathrm{M}_\odot}\right)^{1/2}
\left(\frac{m_{\mathrm{pl}}}{\mathrm{M}_\odot}\right)^{-1}
\left(\frac{a}{1\mathrm{AU}}\right)^{-5/2}\mathrm{yr}.
\end{eqnarray}

Orbit crossing begins more quickly if the perturbing planet is more massive or eccentric, and the time-scale is a strong function of planetesimal semi-major axis.

Equation (\ref{eq:tcrossext}) is in good agreement with equation (14) of \cite{2006Icar..183..193T}, which was derived as an empirical law based on a slightly modified secular theory and N-body simulations, for test particles on circumprimary orbits in a binary system (i.e., external very massive perturber). The chief difference between their result and ours is for high eccentricity.

\section{Relative velocities}\label{s:vrel}

Now that we have established the time-scale on which orbits cross, we proceed to examine the relative velocities of planetesimals undergoing collisions. We shall examine how the distribution of relative velocities evolves with time.

For planetesimals with randomised, uniformly distributed apsides, we might expect the mean relative velocity in a collision to be given by
\begin{equation}\label{eq:evkep}
\left< v_{\mathrm{rel}}\right> = c\left<e\right>v_{\mathrm{kep}},
\end{equation}
where $c$ is a constant of order unity\footnote{For inclined orbits the inclination also plays a role.}. The value of $c$ depends on the specific definition of relative velocity being used, and the underlying eccentricity distribution (see \citealt{1993prpl.conf.1061L}); in our case we wish to know the mean velocity of a planetesimal relative to others in the swarm, for which $c=\sqrt{5/4}$ when the planetesimals' eccentricity follows a Rayleigh distribution, which arises from the mutual gravitational scattering of planetesimals \citep[e.g.,][]{1992Icar...96..107I}. However, the eccentricity distribution arising from a planet's secular perturbations cannot be assumed to be Rayleigh, because the physical process exciting the eccentricities is very different (long-range secular perturbations vs. mutual gravitational scattering). Furthermore, the apsides are constrained by $\varpi \in (\varpi_{\mathrm{pl}}-\frac{1}{2}\pi,\varpi_{\mathrm{pl}}+\frac{1}{2}\pi)$. This is because, with the orbits initially circular, the complex eccentricity starts at the origin of the complex plane and precesses in a circle around $z_{\mathrm{f}}$. Thus it is restricted to the half-plane containing $z_{\mathrm{f}}$. Finally, Equation~(\ref{eq:evkep}) only applies locally, whereas in reality a planetesimal on an eccentric orbit can collide with others over a range of semi-major axes \citep[see][ for a discussion]{2003Icar..162...27T}. In this section, we therefore examine the velocity distribution imposed by planetary secular perturbations, beginning with an estimate for the range of semi-major axes over which collisions can occur.

\begin{figure}
    \includegraphics[width=.45\textwidth]{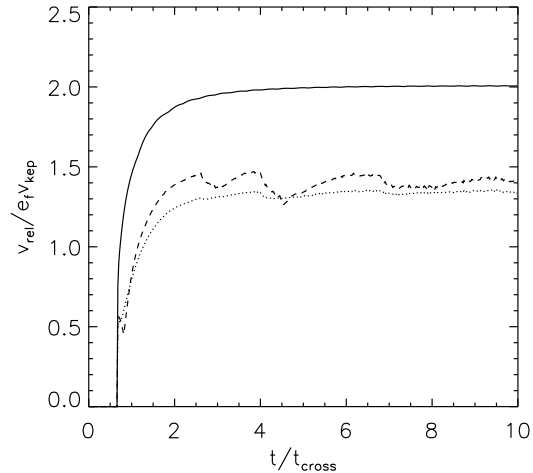}
    \caption{Maximum (solid line), mean (dotted) and median (dashed) relative velocity experienced by a planetesimal, as a function of time. The planetesimal is at 15\,AU, and the planet at 5\,AU with eccentricity $e_{\mathrm{pl}}=0.1$ and mass $m_{\mathrm{pl}}=10^{-3}\mathrm{M}_\odot$, orbiting a solar mass star.}
    \label{fig:vrel}
\end{figure}

First, we estimate the maximum radial excursions of planetesimals evolving under secular perturbations. Consider Planetesimal 1 located at semi-major axis $a_1$, with eccentricity $e_1=\frac{5}{2}a_1e_{\mathrm{pl}}/a_{\mathrm{pl}}$, the maximum attainable under the planet's secular perturbations starting from an initially circular orbit. We wish to find the greatest semi-major axis of an exterior planetesimal, Planetesimal 2, such that Planetesimal 2's orbit can intersect that of Planetesimal 1. If longitudes of periapse could take any angle, this would occur when Planetesimal 1 was at apapse and Planetesimal 2 at periapse, with the orbits tangent, and the apsides antialigned.However, the secular solution also imposes a restriction on the longitude of periapse $\varpi$, restricting it to the range $\varpi \in \left(\varpi_{\mathrm{pl}}-\pi/2, \varpi_{\mathrm{pl}}+\pi/2 \right)$. Because of this, the maximum semi-major axis for Planetesimal 2 must come when Planetesimal 2's orbit is at its lowest eccentricity, i.e., $e_2=0$. Denoting the difference between the semi-major axes of the orbits by $\Delta a$, we find, to lowest order in $e_{\mathrm{pl}}$ and $\alpha$, that the maximum separation of intersecting orbits is
\begin{equation}\label{eq:excursion align}
(\Delta a)_{\max} = \frac{5}{2}a_1\alpha e_{\mathrm{pl}},
\end{equation}
which is simply the maximum radial excursion of Planetesimal 1.

We now numerically calculate the distribution of relative velocities which a planetesimal at $a_1$ experiences, as a function of time, assuming that all the planetesimals evolve deterministically under the Laplace--Lagrange secular solution described in \S\ref{sec:dynamics}, starting on initially circular orbits. For interactions with planetesimals at different semi-major axes, the relative velocity is calculated using formulae in \cite{1998Icar..132..196W}. Note that two confocal ellipses can have two intersection points; we calculate the relative velocity at each point and use both in the analysis below.

Figure~(\ref{fig:vrel}) shows the relative velocities encountered for a fiducial case of $a_{\mathrm{pl}}=5$\,AU, $a_1=15$\,AU, $e_{\mathrm{pl}}=0.1$, $m_\star=1\mathrm{\,M}_\odot$, and $m_{\mathrm{pl}}=10^{-3}\mathrm{M}_\odot$, for $t=0$ to $t=10t_{\mathrm{cross}}$. The surface density of planetesimals is assumed to be constant. We also calculated the averages with a $\Sigma\propto a^{-3/2}$ profile with negligible difference in the derived relative velocities. We see that planetesimals begin on non-intersecting orbits, and evolve for $t\approx t_{\mathrm{cross}}$ before orbits begin intersecting. For these parameters, orbit-crossing begins slightly sooner than $t=t_{\mathrm{cross}}$ because $t_{\mathrm{cross}}$ was derived in the limit $\alpha \ll 1$, and here we have $\alpha=1/3$. We also see that the maximum relative velocity experienced by a planetesimal rises to $\approx 2e_{\mathrm{f}}v_{\mathrm{kep}}$. This is the relative velocity of a planetesimal with eccentricity $2e_{\mathrm{f}}$ relative to a circular orbit \citep{1993prpl.conf.1061L}, but can be achieved for other configurations too. The average relative velocity -- mean or median -- is $\approx 1.4e_{\mathrm{f}}v_{\mathrm{kep}}$. Therefore, we can in fact use Equation~(\ref{eq:evkep}), with $c=2$ if we wish to use the maximum relative velocity, and $c=1.4$ if we wish to use the average. We also note that, while the analytical orbit-crossing criterion only guarantees intersection of infinitesimally separated orbits, we found that soon after orbit crossing began, the planetesimal's orbit intersected the orbits of planetesimals over the whole range permitted by Equation~(\ref{eq:excursion align}).

We can now derive expressions for the maximum relative velocity imposed on a disc by a planet. Combining Equations~(\ref{eq:ef approx}) and (\ref{eq:evkep}) with the expression for the Keplerian velocity $v_{\mathrm{kep}}=\sqrt{\mathcal{G}m_\star/a}$, we get the following expressions for the maximum relative velocities:
\begin{eqnarray}
\max v_{\mathrm{rel}}&\approx& 0.24
\left(\frac{e_{\mathrm{pl}}}{0.1}\right)
\left(\frac{m_\star}{\mathrm{M}_\odot}\right)^{1/2}\nonumber\\
&&\times\left(\frac{a}{10\mathrm{\,AU}}\right)^{-3/2}
\left(\frac{a_{\mathrm{pl}}}{1\mathrm{\,AU}}\right)
\mathrm{km\,s}^{-1}\label{eq:vscaleint}
\end{eqnarray}
for an internal perturber, and
\begin{eqnarray}
\max v_{\mathrm{rel}}&\approx& 0.74
\left(\frac{e_{\mathrm{pl}}}{0.1}\right)
\left(\frac{m_\star}{\mathrm{M}_\odot}\right)^{1/2}\nonumber\\
&&\times\left(\frac{a}{1\mathrm{\,AU}}\right)^{1/2}
\left(\frac{a_{\mathrm{pl}}}{10\mathrm{\,AU}}\right)^{-1}
\mathrm{km\,s}^{-1}\label{eq:vscaleext}
\end{eqnarray}
for an external perturber. If it is desired to work with the mean or median relative velocity, these numbers should be reduced by a factor of $\approx 0.7$.

\begin{figure}
  \includegraphics[width=.5\textwidth]{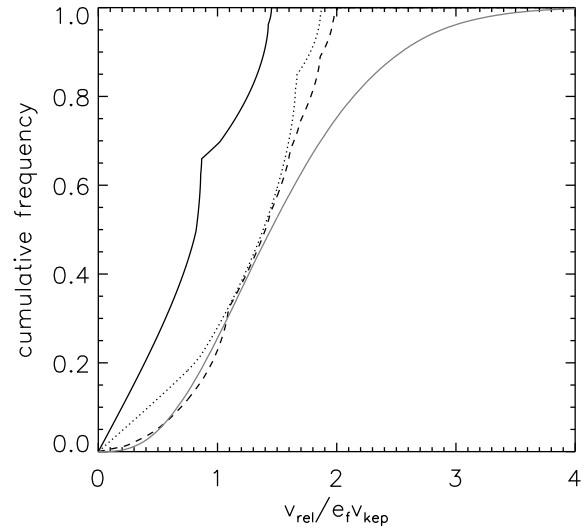}
  \caption{Black: cumulative frequency distribution for relative velocities in a planet-stirred population at $t=1t_{\mathrm{cross}}$ (solid line), $t=2t_{\mathrm{cross}}$ (dotted), and $t=10t_{\mathrm{cross}}$ (dashed). Grey: cumulative frequency distribution for relative velocities in a population with a Rayleigh distribution of eccentricities. The mean eccentricity for the Rayleigh distribution is the same as that for the planet-stirred case at $t=10t_{\mathrm{cross}}$.}
  \label{fig:vcum}
\end{figure}

In Figure~(\ref{fig:vcum}) we compare the relative velocity distribution imposed by a planet's secular perturbations with that arising from a Rayleigh distribution of eccentricities. We see a clear difference in that the Rayleigh distribution gives a tail of high-velocity collisions, in contrast to the planet-stirred distribution's well-defined maximum. In the remainder of this paper, we shall use the maximum relative velocity.

\section{Collision outcomes}\label{s:collision}

We now need to relate the relative velocities in the disc to the outcomes of collisions between planetesimals. When relative velocities are low, collisions between planetesimals result in net growth of the larger body. When relative velocities are high, collisions result in the bodies fragmenting. In the former regime, smaller bodies build up to eventually produce planetary embryos. In the latter regime, planetesimals are ground down in a collisional cascade with the production of large quantities of dust which can be observed as a debris disc. We now proceed to quantify the minimum velocities needed for these erosive collisions to occur, and to find which combinations of planetary parameters can lead to such velocities.

Consider two bodies colliding. The outcome of the collision depends on a large number of factors including the bodies' composition, shape etc. \citep[see, e.g.,][]{1990Icar...84..226H}, but for our purposes the most important considerations are the relative velocity and the kinetic energy of the impact.

We are interested in the transition from net accretion to net erosion, and so use the condition $Q_{\mathrm{R}} > Q^*_{\mathrm{RD}}$, where
\begin{eqnarray}\label{eq:qstarred}
Q_{\mathrm{RD}}^*&=&0.095\mathrm{\,J\,kg}^{-1}\left(\frac{R_{\mathrm{C}1}}{100\mathrm{\,m}}\right)^{-0.33}
\left(\frac{v_{\mathrm{rel}}}{1\mathrm{\,m\,s}^{-1}}\right)^{0.8}\nonumber\\
&& + 0.025\mathrm{\,J\,kg}^{-1}\left(\frac{R_{\mathrm{C}1}}{100\mathrm{\,m}}\right)^{1.2}
\left(\frac{v_{\mathrm{rel}}}{1\mathrm{\,m\,s}^{-1}}\right)^{0.8}
\end{eqnarray}
for weak aggregates \citep{2009ApJ...691L.133S}. Here, $R_{\mathrm{C1}}$ is the radius of a sphere with mass equal to the total of the two bodies' masses $m_1+m_2$, and a density of 1\,g\,cm$^{-3}$, and $Q_{\mathrm{R}}=0.5v_{\mathrm{rel}}^2m_1m_2/(m_1+m_2)^2$ is the reduced mass specific kinetic energy. We use the results for weak aggregates rather than strong rock because firstly we are primarily interested in outer discs which have a high proportion of ices, and secondly we are considering bodies formed by sequential accretion of smaller ones, which results in porous aggregates rather than monoliths \citep{1997Icar..127..290W}.

Here we define the catastrophic disruption threshold velocity, $v_{\mathrm{rel}}^*$, to be the velocity above which collisions between equal-sized bodies no longer result in one of these bodies gaining mass. It is given by
\begin{equation}\label{eq:vrelstar}
v_{\mathrm{rel}}^*(R)=\left[0.8\left(\frac{R}{80\mathrm{\,m}}\right)^{-0.33}
+0.2\left(\frac{R}{80\mathrm{\,m}}\right)^{1.2}\right]^{0.83}\mathrm{m\,s}^{-1},
\end{equation}
where we have converted $R_{\mathrm{C1}}$ into a physical radius $R$ for equal-sized bodies at a density of 1\,g\,cm$^{-3}$. This function has a minimum of $\approx 1$\,m\,s$^{-1}$ at $R\approx 80$\,m.

Equation~(\ref{eq:vrelstar}) gives a lower limit to the relative velocity needed to destroy a body of radius $R$. Because the velocity distribution excited by a planet's secular perturbations has a definite maximum, which decreases with an outer planetesimal belt's semi-major axis, we see that a planet's secular perturbations will be unable to cause catastrophic collisions in a disc beyond some critical semi-major axis
\begin{eqnarray}\label{eq:a*int}
a^*(R)&=&3.8\mathrm{\,AU}
\left(\frac{e_{\mathrm{pl}}}{0.1}\right)^{2/3}
\left(\frac{m_\star}{1\mathrm{M}_\odot}\right)^{1/3}\nonumber\\
&&\times\left(\frac{a_{\mathrm{pl}}}{1\mathrm{\,AU}}\right)^{2/3}
\left(\frac{v_{\mathrm{rel}}^*(R)}{1\mathrm{\,km\,s}^{-1}}\right)^{-2/3}
\end{eqnarray}
for an internal perturber, and within a critical semi-major axis
\begin{eqnarray}\label{eq:a*ext}
a^*(R)&=&1.8\mathrm{\,AU}
\left(\frac{e_{\mathrm{pl}}}{0.1}\right)^{-2}
\left(\frac{m_\star}{1\mathrm{M}_\odot}\right)^{-1}\nonumber\\
&&\times\left(\frac{a_{\mathrm{pl}}}{10\mathrm{\,AU}}\right)^2
\left(\frac{v_{\mathrm{rel}}^*(R)}{1\mathrm{\,km\,s}^{-1}}\right)^2
\end{eqnarray}
for an external perturber.

We can also estimate the size of the largest body which a given planet can destroy at a given disc radius $a$. For $R \gg 80$\,m, we approximate Equation~(\ref{eq:vrelstar}) by $v_{\mathrm{rel}}^* \approx 0.26 (R/80\mathrm{\,m})\mathrm{\,m\,s}^{-1}$; combining this with Equation~(\ref{eq:vscaleint}) for the maximum relative velocity gives
\begin{eqnarray}
R_{\max}&\approx& 74\mathrm{\,km}
\left(\frac{e_{\mathrm{pl}}}{0.1}\right)
\left(\frac{m_\star}{\mathrm{M}_\odot}\right)^{1/2} \nonumber\\
&&\times\left(\frac{a}{10\mathrm{\,AU}}\right)^{-3/2}
\left(\frac{a_{\mathrm{pl}}}{1\mathrm{\,AU}}\right) \label{eq:Rmax}
\end{eqnarray}
for an internal perturber.

\section{Discussion}

\subsection{Ability of a planet to stir a disc} \label{s:effects}

We can now discuss the effects of planetary secular perturbations on an initially dynamically cold planetesimal disc. We have seen how the perturbations induce orbit-crossing (\S\ref{s:timescale}) and excite relative velocities between planetesimals (\S\ref{s:vrel}). Whether this is sufficient to stir a disc depends on the typical size of bodies in the disc (see Equations~\ref{eq:a*int} and \ref{eq:a*ext}). If the planetesimals in the disc have grown to $\sim 80$\,m in size, then a planetary perturber can stir the disc at radii
\begin{equation}\label{eq:a*weak}
a \lesssim a^* = 1800\mathrm{\,AU\,\,}e_{\mathrm{pl}}^{2/3}
\left(\frac{m_\star}{\mathrm{M}_\odot}\right)^{1/3}\left(\frac{a_{\mathrm{pl}}}{1\mathrm{\,AU}}\right)^{2/3}.
\end{equation}
Because the 80\,m bodies are the weakest, Equation~(\ref{eq:a*weak}) determines the greatest range of a planet's secular perturbations for disc stirring. If $R \gg 80$\,m, we can write
\begin{equation}\label{eq:a*general}
a^* = 170\mathrm{\,AU\,\,}e_{\mathrm{pl}}^{2/3}\left(\frac{R}{10\mathrm{\,km}}\right)^{-2/3}
\left(\frac{m_\star}{\mathrm{M}_\odot}\right)^{1/3}\left(\frac{a_{\mathrm{pl}}}{1\mathrm{\,AU}}\right)^{2/3}.
\end{equation}

\begin{figure}
\includegraphics[width=.5\textwidth]{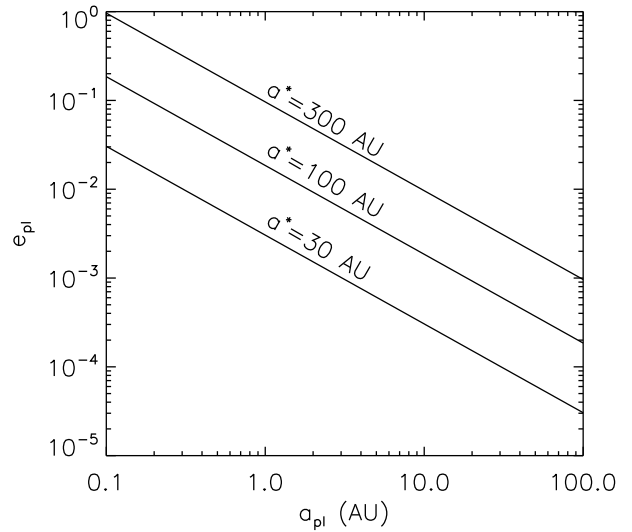}
  \caption{Maximum semi-major axis $a^*$ at which 80\,m bodies can be destroyed, as a function of planetary semi-major axis and eccentricity (Equation~\ref{eq:a*weak}). This is independent of planet mass. Stellar mass $m_\star=2\mathrm{M}_\odot$.}
  \label{fig:apl-a}
\end{figure}

We illustrate Equation~(\ref{eq:a*weak}) in Figure~(\ref{fig:apl-a}). This figure shows the maximum semi-major axis $a^*$ at which 80\,m bodies can be destroyed as a function of a planet's position in $a$--$e$ space. We see that all but very low eccentricity planets close to their star have the potential to stir discs out to at least 30\,AU. Hot Jupiters are therefore ruled out as potential disc stirrers.

We note that this constraint is independent of the planet mass. The planet mass affects the time for perturbations to act, but not their eventual effects, assuming that the planet's secular perturbations are the only source of dynamical evolution. For low mass planets, the secular timescale may be comparable to the timescale for collisional damping of eccentricity \citep*{2004ARA&A..42..549G}.

\subsection{Comparison to self-stirring models}

In the self-stirred model (see \S1), the disc begins in an unexcited ($e\sim 10^{-4}$) state, composed of sub-km planetesimals. These grow through collisions until they reach sizes similar to Pluto's, at which point their gravitational perturbations stir the disc, causing the disc to brighten. On the assumption that both the self-stirring and planet-stirring models are accurately describing the behaviour of the disc, a key question for disc evolution is which occurs sooner. In this section we compare the time-scales for the two processes to occur.

The time taken to form Pluto-sized bodies at a given radius through core accretion, $t_{\mathrm{Pl}}$, is proportional to the orbital time-scale divided by the disc surface density at that radius. Based on extensive numerical simulations, \cite{2008ApJS..179..451K} find
\begin{equation}\label{eq:t_pl}
t_{\mathrm{Pl}}=145x_{\mathrm{m}}^{-1.15}\left(\frac{a}{80\mathrm{\,AU}}\right)^3\left(\frac{2\mathrm{M}_\odot}{m_\star}\right)^{3/2}\mathrm{Myr}.
\end{equation}
Here, $x_{\mathrm{m}}$ parametrises the disc surface density in such a way as to account for the propensity of more massive stars to have more massive discs: the surface density of disc solids is given by
\begin{equation}\label{eq:sigma}
\Sigma=\Sigma_0x_{\mathrm{m}}(m_\star/\mathrm{M}_\odot)(a/a_0)^{-3/2},
\end{equation}
where $\Sigma_0=0.18$g\,cm$^{-2}$ corresponds roughly to the minimum mass solar nebula density at $a_0=30$\,AU. We assume that if giant planets are present then the planetesimal surface density is not depleted below the primordial value by processes such as planet-planet scattering or planetesimal-driven migration. While depletion would lead to longer self-stirring timescales, such processes may also excite the eccentricities of remaining planetesimals, efficiently stirring the disc at earlier times. Here we concentrate exclusively on the planet's secular perturbations.

\begin{figure}
\includegraphics[width=.5\textwidth]{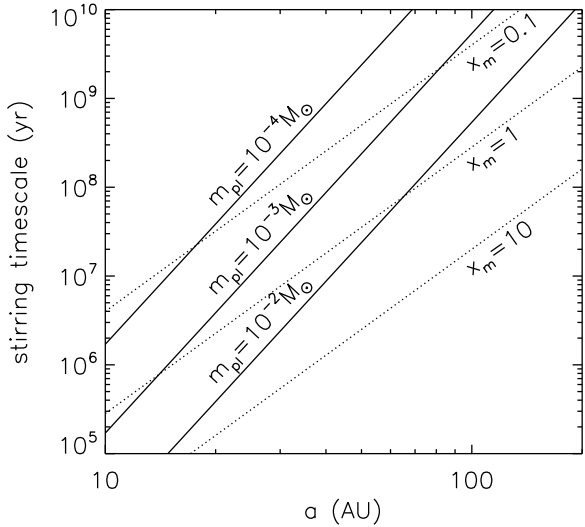}
  \caption{Comparison of the time-scales for stirring in a system with a planet at 5\,AU with eccentricity 0.1, a 2 solar-mass star, and an extended planetesimal disc beyond 10\,AU. Solid lines show planet-stirring time-scales for $m_{\mathrm{pl}}=10^{-4},10^{-3},10^{-2}$\,M$_{\odot}$. Dotted lines show self-stirring time-scales $t_{\mathrm{Pl}}$ for disc masses given by $x_{\mathrm{m}}=0.1,1,10$. More massive planets stir the disc more quickly; more massive discs stir themselves more quickly.}
  \label{fig:times}
\end{figure}

Figure~(\ref{fig:times}) compares the self-stirring time-scale $t_{\mathrm{Pl}}$ given by Equation~(\ref{eq:t_pl}) with the time-scale for planet stirring $t_{\mathrm{cross}}$ given by Equation~(\ref{eq:tcrossint}). At a given planetesimal belt semi-major axis, $t_{\mathrm{Pl}}$ depends only upon the disc surface density, while $t_{\mathrm{cross}}$ depends on planet mass, semi-major axis and eccentricity. Furthermore, both time-scales have a different dependence on stellar mass. As $t_{\mathrm{cross}}$ has a stronger dependence on $a$, for any given planet parameters and disc mass, there exists a radius beyond which the disc will stir itself before the planet can stir it. For example, for a Jupiter-like planet and a disc with $x_{\mathrm{m}}=1$, $t_{\mathrm{cross}} < t_{\mathrm{Pl}}$ out to about 13\,AU, whereas for a heavier 10 Jupiter mass planet $t_{\mathrm{cross}} < t_{\mathrm{Pl}}$ out to around 60\,AU.

With this in mind, we define a new parameter $\Phi$ which describes the boundary between planet-stirred and self-stirred regions of the disc, assuming that planet-stirring and self-stirring are the only two mechanisms operating. The disc is planet stirred if
\begin{eqnarray}\label{eq:phipl}
a \lesssim \Phi=630\mathrm{\,AU\,}\left(1-e_{\mathrm{pl}}^2\right)^{-1}e_{\mathrm{pl}}^{2/3}\left(\frac{m_{\mathrm{pl}}}{\mathrm{M}_\odot}\right)^{2/3}& \nonumber\\
\left(\frac{a_{\mathrm{pl}}}{1\mathrm{\,AU}}\right)^2\left(\frac{m_\star}{\mathrm{M}_\odot}\right)^{-4/3}x_{\mathrm{m}}^{-0.77},&
\end{eqnarray}
and self-stirred otherwise.

\begin{figure}
\includegraphics[width=.5\textwidth]{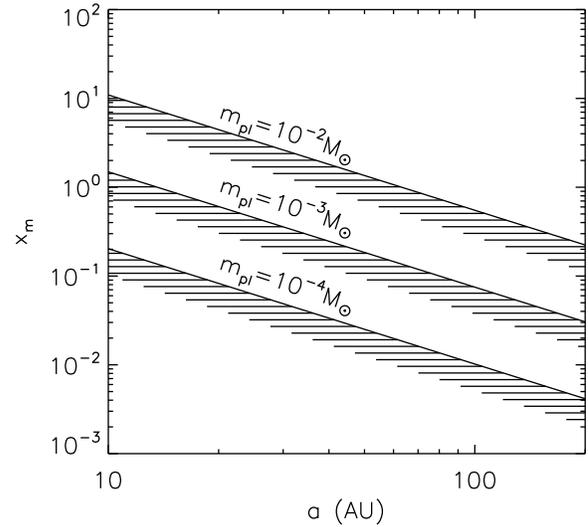}
  \caption{Illustration of constraint (\ref{eq:phipl}), that if planet-stirring is competing with self-stirring then the disc is planet-stirred precisely when the time-scale for planet stirring is shorter than that for self stirring. Lines are plotted at fixed $a_{\mathrm{pl}}=5$\,AU, $e_{\mathrm{pl}}=0.1$, $m_\star=2\mathrm{\,M}_\odot$, and $m_{\mathrm{pl}}$ as shown. At a given disc radius, the planet can stir the disc for disc surface densities on the lower (shaded) side of the line.}
  \label{fig:xm-a}
\end{figure}

This is illustrated in Figure~(\ref{fig:xm-a}), showing the disc surface density below which a planet at 5\,AU, with $e_{\mathrm{pl}}=0.1$,  can stir faster than the disc can form Plutos. If the planet is of Jupiter mass, it can stir discs with $x_{\mathrm{m}}<1$ at 20\,AU.

If the disc is planet-stirred before it self-stirs, then there will be implications for planet formation. If the largest bodies that the planet can destroy (Equation~\ref{eq:Rmax}) are larger than the largest bodies in the disc, then further growth of planetesimals will be difficult if not impossible. Even if the largest planetesimals are unable to be destroyed, their growth rates will be reduced as the increased velocity dispersion amongst the planetesimals reduces gravitational focusing factors, leading to longer collision time-scales.

\subsubsection{Observables of stirring models}

If the disc has an inner hole then we might expect a low level of dust production, hence IR luminosity, until the planet's secular perturbations cause orbits to cross at the inner disc edge. The disc would then brighten (at a given wavelength), before getting dimmer again at that wavelength as the region of peak dust production moves outwards. This could explain the observed incidence of excess IR emission and the fractional luminosity for young A-type stars, which both apparently increase with age, peaking at 10--20\,Myr before declining \citep{2008ApJ...688..597C}. This behaviour is qualitatively similar to that predicted by self-stirring, where the region of peak dust production moves outwards as Pluto-sized bodies form at progressively larger radii. The peak at 20\,Myr cannot therefore be taken as evidence for one particular type of delayed stirring over another.

If most planetesimal discs are extended rather than being narrow rings, we would expect to see an increase in observed disc radius if this is tracing where the disc has recently been stirred. While in principle this would provide a way to discriminate between different stirring mechanisms (for self-stirring $r\propto t^{1/3}$ while for planet-stirring $r\propto t^{2/9}$), the difference is so small that it would be very difficult to distinguish in practice. In any event, there is not currently evidence for any dependence of disc radius on time \citep{2005ApJ...635..625N}.

\subsubsection{Discs unlikely to be self-stirred}

Some stars are young ($\sim 10$\,Myr) yet already have bright debris discs of large radius ($\sim 100$\,AU). Such discs cannot have self-stirred unless the disc is sufficiently dense, and in such systems planet-stirring may be a viable alternative. To quantify this, we calculated the minimum $x_{\mathrm{m}}$ required for a disc to self stir in less than the system age, for discs around 23 FGK stars and 35 A stars with published 24/25 and 70/60 micron excesses \citep{2006ApJ...644..525M,2006ApJ...652.1674B,2006ApJ...653..675S,2007ApJ...658.1289T, 2008ApJ...677..630H}. Disc radii were estimated by fitting black-body curves to the IR excess. For FGK stars we increased these radii by a factor of three because a comparison with the radii known directly from those discs which have been imaged showed that the black body fits systematically underestimate the radii by roughly this amount; this is likely due to the small blow-out size for dust in these discs. If radii were available from imaging, we used these in preference to the black-body fits. We identified that a disc would have trouble self-stirring if $x_{\mathrm{m,min}} \ge 10$.

Very massive discs would also have been gravitationally unstable when gas was still present. We can calculate the minimum density for gravitational instability using the Toomre criterion $Q=\frac{c_{\mathrm{s}}n}{\pi \mathcal{G}\Sigma_{\mathrm{g}}}\lesssim 1$ for instability, where $n$ is the mean motion of the disc, $c_{\mathrm{s}}$ is the sound speed in the disc gas, and $\Sigma_{\mathrm{g}}$ is the surface density of gas. The sound speed $c_{\mathrm{s}}\approx (h/a)v_{\mathrm{kep}}$ \citep{1981ARA&A..19..137P}, where $h$ is the disc scale height. Assuming $h/a=0.1$ and a dust:gas ratio of 1:100, so that $\Sigma_{\mathrm{g}}=100\Sigma$ with $\Sigma$ given by Equation~(\ref{eq:sigma}), this gives
\begin{equation}\label{eq:toomre}
Q\lesssim 10^2 \left(\frac{a}{1\mathrm{\,AU}}\right)^{-1/2}x_{\mathrm{m}}^{-1}
\end{equation}
for instability. For a disc at 100\,AU, this corresponds to a maximum $x_{\mathrm{m}}$ of $\sim 10$. Higher $x_{\mathrm{m}}$ would still be possible through metallicity enhancement without affecting the gas mass and therefore gravitational stability.

\begin{figure}
\includegraphics[width=.5\textwidth]{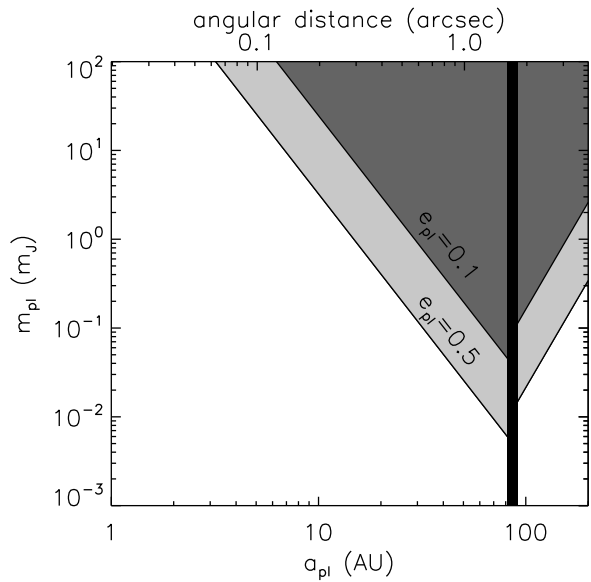}
  \caption{The shaded region shows the region of $a_{\mathrm{pl}}$--$m_{\mathrm{pl}}$ parameter space a perturbing planet must occupy to stir the disc in the HD~181327 system in less than the system age of $\sim$12\,Myr, assuming a planetary eccentricity of $0.1$ or $0.5$. Planets with higher eccentricity can stir the disc if they have lower mass or semi-major axis than ones with lower eccentricity. The disc is indicated with the thick vertical line.}
  \label{fig:hd181327}
\end{figure}

We identify two discs with a minimum surface density for self-stirring $x_{\mathrm{m,min}} \ge 10$: HD~181327 ($x_{\mathrm{m,min}} = 17$) and HD~202917 ($x_{\mathrm{m,min}} = 10$). Both these discs have been imaged, with radii $a=86$\,AU \citep{2006ApJ...650..414S} and $a\approx 80$\,AU \citep{2007lyot.confE..32K} respectively. It may be that such discs do indeed have $x_{\mathrm{m,min}} \ge 10$; i.e., they may be at the top end of the disc mass distribution, in which case they may be self-stirred, assuming that they have managed to avoid the gravitational instability mentioned in the previous paragraph. In the case of HD~181327, however, there is independent evidence in support of planet stirring: the disc has an azimuthal asymmetry \citep{2008ApJ...689..539C} which could be due to planetary secular perturbations \citep{1999ApJ...527..918W}, so this system in particular warrants further investigation. In such a disc, we can place constraints on the parameters of the planet responsible for stirring by requiring $t_{\mathrm{cross}} < t_{\mathrm{age}}$. Figure~\ref{fig:hd181327} shows that such planets are likely to be be far from the star. This, together with the host stars' youth, makes them good targets for direct imaging.

\subsection{Exoplanet population}

We now attempt to ascertain whether there is any observational evidence for planet stirring. We begin by looking at a statistical sample of exoplanets. If planets are a common cause of disc stirring then we might expect there to be a correlation between the planetary parameters $\Phi$ and $a^*$ and the presence of infrared excess, higher values of these parameters correlating with IR excess. Of the two, $a^*$ is the more fundamental because it describes the planet's absolute ability to stir a disc within the context of the planet-stirring model, independently of any other sources of stirring. A star hosting a planet may also host a planetesimal disc. If it does, and if planet-stirring were the sole stirring mechanism, then we would expect only those planets with high enough $a^*$ to exhibit IR excess. The parameter $\Phi$ quantifies the relative importance of the planet- and self-stirring models, so the interpretation of any correlation between high $\Phi$ and a disc, should one exist, is not so clear.

We take 57 planet-hosting stars with published \emph{Spitzer/MIPS} photometry \citep{2006ApJ...652.1674B,2007ApJ...658.1312M,2008ApJ...674.1086T,2009ApJ...690.1522B}. Our sample is identical to that of Bryden et al. (in prep.), but with the two M dwarfs GJ~436 and GJ~876, and the G dwarf HD~33636 whose companion has been determined to be of stellar mass \citep{2007AJ....134..749B}, removed. Of these 57 stars, 10 show significant excess emission at 24 $\mu$m and/or 70 $\mu$m, and are classed as disc hosts. Exoplanet data are from \cite{2006ApJ...646..505B} for most planets, except for $\epsilon$~Eridani \citep{2006AJ....132.2206B} and HD~69830 \citep{2006Natur.441..305L}.

\begin{figure}
\includegraphics[width=.5\textwidth]{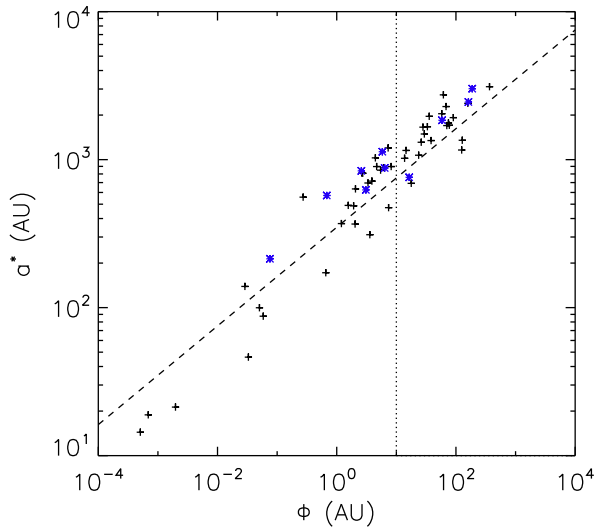}
  \caption{Plot of $a^*$ against $\Phi$ for systems in the sample of Bryden et al. (in prep.). We plot only the planet which stirs the disc quickest for multi-planet systems. Black crosses show planets orbiting stars without discs. Blue stars show planets orbiting stars with discs. Dotted line marks 10\,AU. Dashed line marks $a^*$ against $\Phi$ for planets with $e_{\mathrm{pl}}=0.1$, $m_{\mathrm{pl}}=10^{-3}\mathrm{\,M}_\odot$, and varying $a_{\mathrm{pl}}$. We set $x_{\mathrm{m}}=1$.}
  \label{fig:bryden}
\end{figure}

In Figure~(\ref{fig:bryden}) we plot parameters $\Phi$ and $a^*$ for the 57 systems. For multi-planet systems, we plot $\Phi$ and $a^*$ for the planet which stirs the disc quickest, treating the system as if the planets' perturbations acted independently; see \S\ref{s:multi} for a more detailed discussion. We also plot the line for constant $e_{\mathrm{pl}}=0.1$, $m_{\mathrm{pl}}=10^{-3}\mathrm{\,M}_\odot$ (a `typical planet'), and varying $a_{\mathrm{pl}}$, which fits the points rather well, showing that most of the range of $\Phi$ and $a^*$ can be accounted for by spread in $a_{\mathrm{pl}}$. The planet's semi-major axis is the most important parameter in determining whether a disc can be self- or planet-stirred. The region on the right shows the region of parameter space in which a planet can stir an $a=10$\,AU, $x_{\mathrm{m}}=1$ disc, according to Equations~(\ref{eq:phipl}) and (\ref{eq:a*weak}).

Figure~(\ref{fig:bryden}) does not suggest a difference in the distributions of $\Phi$ and $a^*$ between planets orbiting disc hosting and non disc hosting stars. This is confirmed by Kolmogorov--Smirnov tests. The $p$-values\footnote{The $p$-value gives the probability of observing a more extreme test statistic under the assumption of the null hypothesis: that the distribution of the parameter is the same for both samples. Small $p$-values suggest that the populations have different distributions.} from one-dimensional KS tests comparing the disc hosting and non disc hosting samples are 0.986 when comparing the distributions of $\Phi$ and 0.917 when comparing the distributions of $a^*$. Thus the distributions of planetary parameters are statistically indistinguishable between disc hosts and non disc hosts. 

The implications of this for the relative importance of self- and planet-stirring are however unclear. Due to the many processes doubtless taking place, it is likely that should any correlation be present it has been diluted. Larger samples, at a range of ages (all but two of the stars in this sample are over 1\,Gyr old), may be necessary to properly determine the evolutionary processes at work.

\subsection{Case studies} \label{s:case}

We now proceed to examine some individual systems, categorising them somewhat arbitrarily by the planet's semi-major axis. We conclude this subsection by briefly looking at multi-planet systems.

\subsubsection{Jupiter analogues}

\begin{figure}
\includegraphics[width=.5\textwidth]{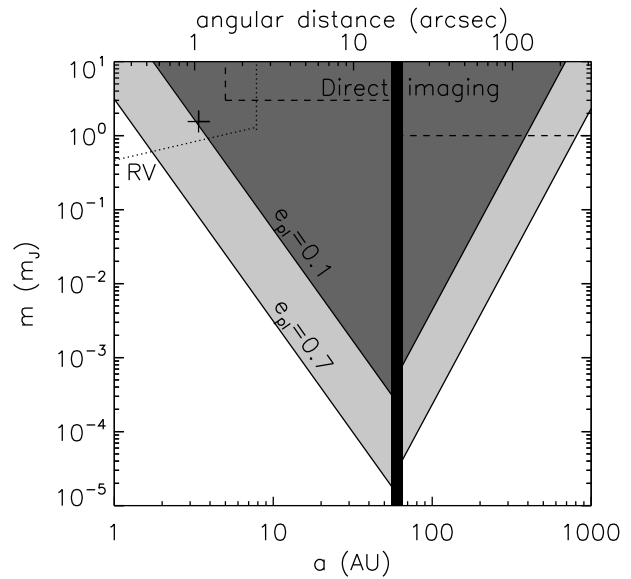}
  \caption{Planet parameters needed to stir $\epsilon$ Eridani's disc at 60\,AU, within the age of the system. The diagonal lines show the minimum planet mass at each semi-major axis required to stir the disc for $e_{\mathrm{pl}}=0.1$ and $e_{\mathrm{pl}}=0.7$. We also show detection limits for RV (dotted line) and direct imaging (dashed line). The real location of $\epsilon$ Eridani~b is marked with a cross.}
  \label{fig:epseri}
\end{figure}

$\epsilon$~Eridani hosts a highly eccentric ($e_{\mathrm{pl}}=0.7$) Jupiter-mass ($m_{\mathrm{pl}}=1.5\times 10^{-3}\mathrm{M}_\odot$) planet at 3.4\,AU \citep{2006AJ....132.2206B}. This planet can stir any disc of 80\,m planetesimals out to $a^*=3000$\,AU. $\epsilon$~Eridani also hosts a cold debris disc extending from 35--110\,AU, with surface brightness peaking at 60\,AU, as well as at least one unresolved warm inner belt \citep{2005ApJ...619L.187G,2009ApJ...690.1522B}. Clearly, all the dust is well within the maximum radius for planet-stirring by the $a^*$ criterion. We can also compare $t_{\mathrm{cross}}$ to the star's age, estimated at $\sim 850$\,Myr \citep{2004A&A...426..601D}. Because $t_{\mathrm{cross}}\approx 40$\,Myr at 60\,AU, we see that planet-stirring by planet b is inevitable within the system age. Figure~(\ref{fig:epseri}) shows the region of parameter space within which a planet must lie if it is to have stirred the disc within 850\,Myr, for both a fiducial planetary eccentricity $e_{\mathrm{pl}}=0.1$ and the real planetary eccentricity $e_{\mathrm{pl}}=0.7$. The figure also shows the parameter space accessible to 20 years' radial velocity observations at 15\,m\,s$^{-1}$ precision, and the rough sensitivity limits from the direct imaging searches of \cite{2006ApJ...647.1437M} and \cite{2008A&A...488..771J}.

We can also calculate the maximum size of bodies that can be destroyed by planet stirring by $\epsilon$~Eridani~b. At 60\,AU, Equation~(\ref{eq:Rmax}) gives $R_{\max}\approx 110$\,km. The planetary perturbations therefore have an impact over a wide range of the size distribution of planetesimals.

It is intriguing that the planet $\epsilon$ Eridani b is only just able to stir the disc at 110\,AU, within 850\,Myr. This may be coincidence, but may hint that there is an as yet unstirred disc region beyond 110\,AU, which the planet's perturbations have not yet reached. It is, however, worth noting that the disc could also be stirred by any other planet in the system, such as that postulated to explain the disc's clumpy structure \citep{2005ApJ...619L.187G}, although here the interactions would not be purely secular.

\subsubsection{Neptune analogues} \label{s:Neptunes}

Fomalhaut's long-suspected planet was recently imaged by \cite{2008Sci...322.1345K}. Its orbital elements are estimated at $e_{\mathrm{pl}}=0.11$, and $a_{\mathrm{pl}}=115$\,AU, with an upper limit of $3\mathrm{\,M}_{\mathrm{J}}$ for the mass \citep{2009ApJ...693..734C}. The disc lies at a radius of $\sim 140$\,AU \citep*{2005Natur.435.1067K}. At this radius, the time-scale for planet stirring is only 0.65\,Myr, orders of magnitude less than the star's age ($\sim 200$\,Myr, \citealt{1997ApJ...475..313B}). We also find $a^*=1.2 \times 10^4$\,AU, so the disc is well within the radial limits for planet-stirring.

\begin{figure}
\includegraphics[width=.5\textwidth]{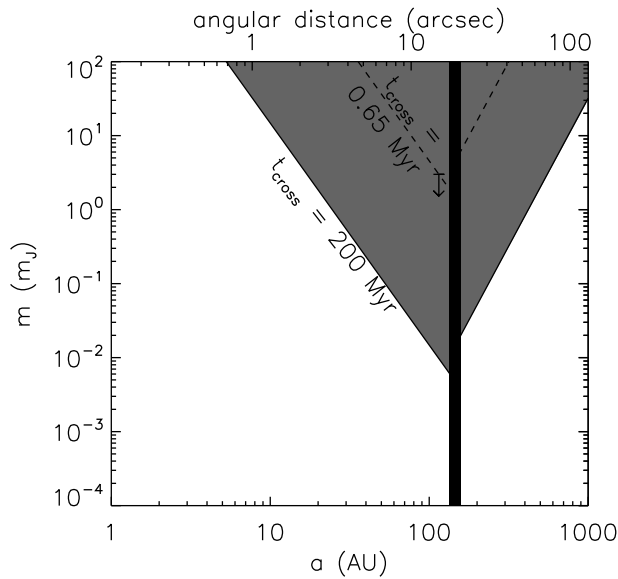}
  \caption{Planet parameters needed to stir the disc of Fomalhaut. The shaded region marks the planet parameters where disc stirring can occur within the system age. The dashed line marks the planet parameters which can stir the disc as quickly as Fom~b (0.65\,Myr). The planet's semi-major axis and maximum mass are marked with an arrow.}
  \label{fig:fom}
\end{figure}

Figure~(\ref{fig:fom}) shows the planet masses and semi-major axes required to stir the Fomalhaut disc within the system's age, assuming a planetary eccentricity of 0.1. We can see that even if the planet's mass is significantly less than the maximum of $\sim 3$\,M$_{\mathrm{J}}$, planet-stirring would occur within the age of the system.

However, Fomalhaut presents two complications. Firstly, there are difficulties with \emph{in situ} formation of the planet Fom b because the time-scales for core accretion are so long: recall that the self-stirring time-scale gives the time required to form \emph{Pluto}-sized objects \emph{in situ}, while the mass of Fom b may be as high as that of Jupiter. The time to form a Pluto-sized body at Fomalhaut b's orbit is around 150\,Myr for $x_{\mathrm{m}}=1$. The planet most likely formed closer to the star and later moved to its current location, for example by outwards migration \citep[e.g.,][]{2007MNRAS.378.1589M} or being scattered by another planet (e.g., \citealt*{2009arXiv0902.2779V}). Both of these processes would however likely disturb the disc as well.

Secondly, we note that, although $e_{\mathrm{f}}\approx 0.1$, the material in the Fomalhaut disc appears to have very low proper eccentricities \citep{2006MNRAS.372L..14Q,2009ApJ...693..734C}, as evidenced by the sharp inner edge to the disc. If the proper eccentricity of Fomalhaut's disc is only 10 per cent of the forced eccentricity then this increases the time-scale for orbit crossing to 6.5\,Myr (see Equation~\ref{eq:tcross}), still much less than the system's age. Reducing proper eccentricities also reduces the relative velocities amongst planetesimals in direct proportion, although given the large value of $a^*$ this will not prevent Fom~b from causing erosive collisions.

For the Solar System's Neptune we find $a^*=730$\,AU, making the Kuiper Belt able to be stirred by Neptune. However, when we compare with self-stirring we find that planet-stirring acts more quickly only out to $\Phi = 33$\,AU, so Neptune's secular perturbations would not have stirred the belt before Pluto formed, assuming that the planets formed at their current semi-major axes\footnote{We find similar values for other giant planets; e.g., for Jupiter, $a^*=720$\,AU and $\Phi=22$\,AU}. So this simple model is consistent with the outer Solar System, although we note that the dynamical evolution of the early Kuiper Belt and outer planets may have been more complicated than \emph{in-situ} formation of Neptune followed by growth of Kuiper Belt Objects \citep{2005Natur.435..459T}. We also note that highly excited eccentricities and inclinations of KBOs may have been required to explain the details of the capture of Neptune's Trojans \citep{2009AJ....137.5003N}, and capture of KBOs into high order mean motion resonances (e.g., \citealt{2003AJ....126..430C}). Such high inclinations might be achievable through self-stirring but not planet-stirring.

\subsubsection{Multi-planet systems} \label{s:multi}

When dealing with multiple planets previously we treated the disc as being stirred by the planet with the lowest $t_{\mathrm{cross}}$, assuming that the other planets had no effect on the disc. Such an approach is unrealistic because it neglects not only the effects of other planets on the disc, but also the mutual interactions of the planets amongst themselves. 

\begin{figure}
\includegraphics[width=.5\textwidth]{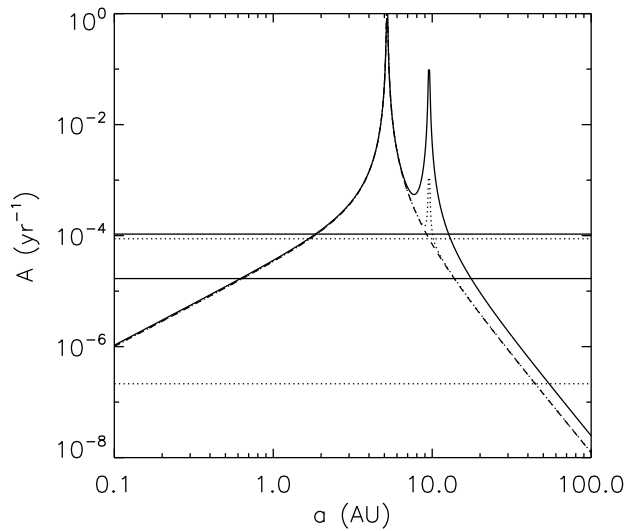}
  \caption{Solid lines: Precession rate $A$ for planetesimals in the Sun-Jupiter-Saturn system. The horizontal lines mark the eigenfrequencies. Dotted lines: precession rate and eigenfrequencies for the same system but with Saturn's mass reduced to that of Earth. Dashed line: precession rate for planetesimals perturbed by Jupiter alone. Secular resonances occur when the precession rate equals one of the system's eigenfrequencies; there are no secular resonances in the case of a single perturber.}
  \label{fig:A-jupsat}
\end{figure}

We plot the precession rate $A$ for planetesimals orbiting in the Sun-Jupiter-Saturn system in Figure~(\ref{fig:A-jupsat}). This also shows the location of secular resonances, where the planetesimal's precession rate equals one of the system's eigenfrequencies and the forced eccentricity is formally infinite.

Figure~(\ref{fig:A-jupsat}) also shows the effect of reducing Saturn's mass to that of Earth: the precession rate approaches that in the single-planet case of Jupiter alone, and the width of the region strongly affected by the outer planet's perturbations decreases. So as far as the precession rate is concerned, the behaviour is similar to the single-planet case.

Performing a similar analysis to that in \S\ref{s:timescale}, we find that, for planetesimals on initially circular orbits, the time-scale for orbit crossing in the multi-planet case is given by
\begin{equation}\label{eq:tcrossmulti}
t_{\mathrm{cross}} \gtrsim \left(\left|\sum\limits_{j=1}^Nb_jB_j\right|\times\sum\limits_{i,j=1}^N\left|\frac{A_je_{ji}}{A-g_i}\right|\right)^{-1},
\end{equation}
where $b_j=7/2$ or $-3/2$ when planet $j$ is an internal or external perturber respectively.

\begin{figure}
\includegraphics[width=.5\textwidth]{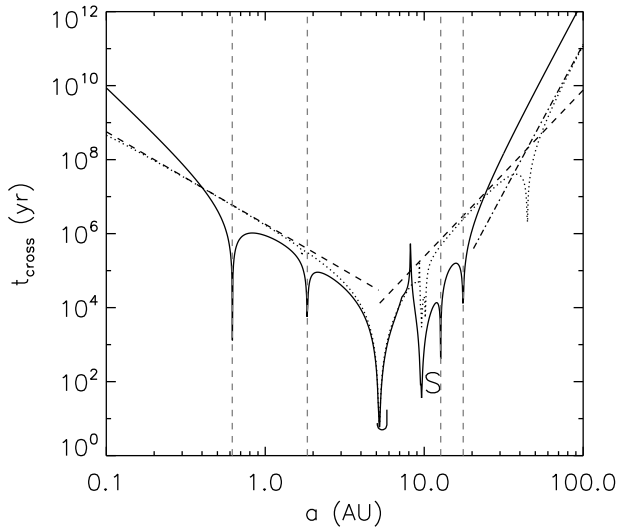}
  \caption{Solid line: orbit-crossing time-scale $t_{\mathrm{cross}}$ for planetesimals in the Sun-Jupiter-Saturn system. Dotted line: same, with Saturn's mass reduced to that of Earth. Dashed lines: $t_{\mathrm{cross}}$ assuming a one-planet system with Jupiter stirring the disc. Dot-dash line: the asymptotic approximation to the solid for small $\alpha$. As well as being short near the planets (marked with their initials), $t_{\mathrm{cross}}$ is also small near the four secular resonances (marked with vertical lines for the Sun-Jupiter-Saturn system).}
  \label{fig:tcross-jupsat}
\end{figure}

The stirring time $t_{\mathrm{cross}}$ for the Sun-Jupiter-Saturn system is plotted in Figure~(\ref{fig:tcross-jupsat}). We also show $t_{\mathrm{cross}}$ with Saturn's mass reduced to that of Earth, and for the single-planet case with Jupiter alone perturbing the disc. With Saturn at its true mass, we see that the crossing timescale is greatly reduced close to the planets. However, beyond the outermost secular resonance, the dependence of $t_{\mathrm{cross}}$ on $a$ steepens. Specifically, for large $a$, $t_{\mathrm{cross}}\propto a^8$ rather than $a^{4.5}$. This is because we now have $|A| < |g_i|$ in the forced eccentricity term in Equation~(\ref{eq:tcrossmulti}), and so the $a$ dependence of $A$ and $B_j$ no longer cancels. For planetesimals beyond 20\,AU, introducing another perturber has \emph{increased} the time-scale for orbit crossing. When Saturn's mass is reduced to that of Earth, we see a large region between Saturn and the outer secular resonance where the the time-scale is the same as for the case with Jupiter alone: because of the large disparity in masses, the perturbations are dominated by Jupiter.

\begin{figure}
\includegraphics[width=.5\textwidth]{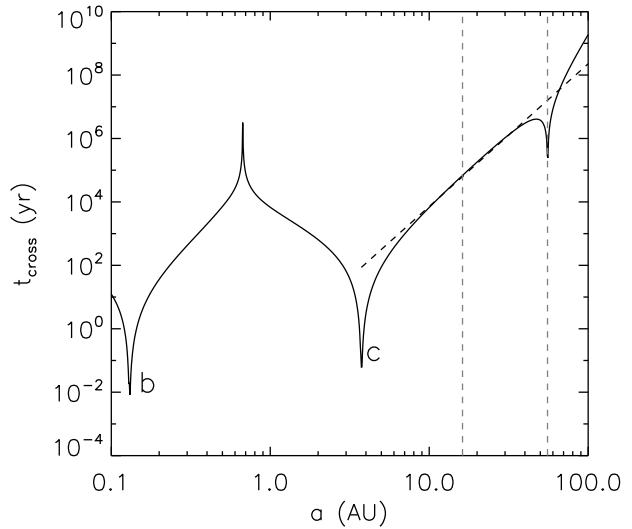}
  \caption{Solid line: orbit-crossing time-scale $t_{\mathrm{cross}}$ for the HD~38529 system. Planets are shown with their letters, and the secular resonances are marked with vertical lines. The dashed line shows the crossing time if only planet c is stirring the disc. The sharp peak at $\approx 0.7$\,AU occurs where $\mathrm{d}A/\mathrm{d}a=0$ (see text).}
  \label{fig:hd38529}
\end{figure}

As an example of a multi-planet system with a debris disc, consider HD~38529. This star hosts a 0.8 Jupiter mass planet on a 0.13\,AU orbit and a 12.2 Jupiter mass planet on a 3.74\,AU orbit. The secular dynamics of the system, including both planets and massless planetesimals, were modelled by \cite{2007ApJ...668.1165M}, who concluded from the dynamical analysis and SED fitting that the planetesimals reside in a dynamically stable region at 20--50\,AU between secular resonances. Figure~(\ref{fig:hd38529}) shows the crossing time-scale for HD~38529. Within the region 20--50\,AU, the crossing time-scale is close to that achieved by planet c alone, due to its higher mass and larger semi-major axis. Within this region of the disc, the planets' secular perturbations induce crossing of neighbouring initially circular orbits on time-scales of $\lesssim 1$\,Myr. It may well be the case then that there are no bodies larger than several kilometers in radius in the disc (but see \S\ref{s:ic}).

Note that $t_{\mathrm{cross}}$ is formally infinite when $\mathrm{d}A/\mathrm{d}a=0$. This would appear to mean that there is a particular semi-major axis between the planets where the perturbations can never induce orbit-crossing. However, this singularity is merely a mathematical artefact (see \S\ref{s:timescale}): in reality, this region can still be stirred by secular perturbations, given sufficient time.

\subsection{Limitations of model}

We have presented a simple picture of the effect of secular perturbations on a planetesimal disc. Here we clarify the assumptions and limitations attached to our model.

\subsubsection{Collision model}

To determine the outcome of collisions, we have chosen one particular scaling law for threshold collision energy. Other scaling laws differ in the planetesimal radius at which the minimum of $Q_{\mathrm{D}}^*$ is attained, and the value of the minimum itself. Given the very large values of $a^*$ for 100\,m bodies, this is unlikely to be important for this size of planetesimal, but may be important if the planetesimals are larger.

We have assumed that collisions occur as soon as the orbits begin to cross, and accounting for collision rates will slightly increase the stirring time.

\subsubsection{Initial conditions} \label{s:ic}

In common with other studies \citep[e.g.,][]{2006Icar..183..193T,2008ApJS..179..451K}, we have taken the initial planetesimal orbits to be circular. Despite promising recent progress \citep[e.g.,][]{2007Natur.448.1022J,2008ApJ...687.1432C}, the formation of planetesimals is still not fully understood, and so these initial conditions, although reasonable, are not rigorously justified. In particular, we note that planetesimals can acquire moderate ($\sim 0.05$) eccentricities if they orbit within a protoplanetary disc that has density fluctuations induced by its self-gravity \citep*{2008MNRAS.385.1067B} or by the MRI \citep{2005A&A...443.1067N}. Starting at different eccentricities can be dealt with by changing the proper eccentricity $e_{\mathrm{p}}$ in Equation~(\ref{eq:tcross}), so long as $e_{\mathrm{p}}$ is a single-valued function of semi-major axis.

We have also introduced the planet instantaneously, assuming that it forms at time $t=0$. Statements about the stirring time-scale should therefore be qualified by including the time taken for the planet to form, during which time the planetesimals will themselves be growing. This will affect the maximum radius of bodies which exist in the disc when the orbits begin to cross.

\subsubsection{Unmodelled processes}

Dynamically, we have neglected any non-secular dynamical effects of the planet on the disc. While this is valid for the planetesimals far from the planet, when they are close they begin to experience resonant interactions as well as secular (see Figure~\ref{fig:tsec-a}). Furthermore, the asymptotic expression for crossing time-scale (Equation~\ref{eq:tcrossint}) is no longer valid. In fact, this approximation over-estimates the time-scale because, as $a \to a_{\mathrm{pl}}$, we find $A\to\infty$ while $e_{\mathrm{f}}\to e_{\mathrm{pl}}$ \citep{1999ssd..book.....M}, and so $t_{\mathrm{cross}}\to 0$, whereas extrapolation of the asymptotic expression predicts a finite value. So both simplifying the secular interactions and neglecting non-secular interactions will tend to overestimate the crossing time-scale. Thus the simplified dynamics underestimates the ability of planets to stir discs located close to them. This may be particularly relevant for such systems as Fomalhaut, where several strong resonances lie in the disc \citep{2009ApJ...693..734C}. 

We have neglected any damping in the disc. Collisional damping has been invoked to explain the low proper eccentricities in Fomalhaut's disc \citep{2006MNRAS.373.1245Q,2009ApJ...693..734C}. Reducing proper eccentricity increases both the time-scale for orbit crossing and the relative velocities, as we have already described.

Finally, while we have focussed on planetary secular perturbations as a disc stirring mechanism, and compared them to \emph{in situ} planet formation, it is important to realise that there may be other causes of dynamical excitation. \cite{2002AJ....123.1757K} investigated the effects of a stellar flyby, but found that the perturbations were rapidly damped by collisions. Other mechanisms which have not been so thoroughly investigated in the context of debris disc evolution include planet formation proceeding more rapidly when a gap-opening giant planet has formed \citep{2000ApJ...540.1091B,2005ApJ...626.1033T}, or an outer disc being stirred by planetesimals that have been scattered out from the inner system \citep{2004ApJ...614..497G}. 

\section{Summary and conclusions}

Our main conclusion is that a planetesimal belt at several 10's of AU can be stirred by an eccentric giant planet at only a few AU. Debris discs do not require any bodies larger than a few km in size beyond a few AU from the star to explain the observed dust production.

To reach this conclusion, we investigated the effects of secular perturbations from an eccentric planet on a dynamically cold disc, to assess whether this might be the origin of debris disc stirring. Over time, neighbouring orbits acquire sufficiently different eccentricities and longitudes of periapse that they begin to cross. We derived an analytical expression for the time $t_{\mathrm{cross}}$ for this to occur (Equation~\ref{eq:tcrossext}), which agrees well with previously published N-body simulations. After this time, the planet's secular perturbations quickly impose a mean relative velocity $\left<v_{rel}\right> \approx 1.4e_{\mathrm{f}}v_{\mathrm{kep}}$. This is similar to that for a planetesimal swarm without external perturbers, despite the high degree of apsidal alignment forced on the orbits by secular perturbations.

When the relative velocities increase, the disc may brighten as a result of increased dust production, and further growth of planetesimals may be inhibited or halted. We derived an expression for $a^*$, the maximum range of semi-major axes over which a planet's perturbations can destroy planetesimals (Equation~\ref{eq:a*weak}). The range $a^*$ increases with planetary eccentricity and strongly increases with planetary semi-major axis. It is also a function of planetesimal size. For the weakest planetesimals ($\sim 100$\,m in radius), we found typical values of $a^*$ of several hundred AU. 

We then compared the time-scales for planet stirring with the time-scales for self stirring from the models of \cite{2008ApJS..179..451K}. Because $t_{\mathrm{cross}}$ has a stronger dependence on $a$ than does $t_{\mathrm{Pl}}$, assuming that only these two proceeses are operating, we found that typically the disc closer to the planet will be planet-stirred, and the disc further away will be self-stirred, and we identified another parameter $\Phi$ which demarcates the outer reach of the planet's perturbations, beyond which the self-stirring time-scale is shorter. This parameter $\Phi$ is typically much smaller than $a^*$, and whether a disc can be planet-stirred before it is self-stirred depends on the disc density as well as planet parameters.

For a sample of RV planet hosts observed by \emph{Spitzer,} we find no correlation between the magnitude of $\Phi$ or $a^*$ and the presence of a disc. While this may suggest that planet stirring is not ubiquitous, the degree of correlation we would expect is not clear.

However, for some individual systems it appears that a known planet will have stirred the disc on a time-scale shorter than the system age and/or before it it likely to self-stir. In particular, we identify $\epsilon$~Eridani, Fomalhaut, and HD~38529 as being in this category. Future studies of such discs should take into account the effects of planetary perturbations on the disc's evolution, collisional as well as dynamical. We also speculate that the discs of HD~181327 and HD~202917 may have been stirred by as yet undiscovered planets, since the only other stirring mechanism proposed would only work for discs with high surface densities ($\ge 10$\,MMSN).

Future work in this area should further investigate the effects of planetary perturbations on the collisional evolution of a disc, with a view to \emph{(a)}~clarifying the extent to which the perturbations can inhibit the further growth of planetesimals and \emph{(b)}~determining the evolution of dust production and hence IR luminosity. The latter in particular will enable valuable observational tests of planet-stirring and its role \emph{vis-a-vis} other stirring mechanisms. We also note that, while we have focused on internal perturbers in our discussion, our formulae for external perturbers will be relevant for investigations into the effects of binary companions on discs.

\section*{Acknowledgments}

AJM is grateful for the support of an STFC studentship. The authors wish to thank Z. Leinhardt for useful discussions on collisions, and the reviewer John Chambers for suggesting a number of improvements to the paper.

\bibliographystyle{/home/ajm233/latex/mnras/mn2e}
\bibliography{bibliography}

\end{document}